\documentclass[conference]{IEEEtran}
\IEEEoverridecommandlockouts
\usepackage{cite}
\usepackage{amsmath,amssymb,amsfonts}
\usepackage{algorithmic}
\usepackage{graphicx}
\usepackage{textcomp}
\usepackage{xcolor}
\usepackage{hyperref}
\DeclareUnicodeCharacter{202B}{ }
\DeclareUnicodeCharacter{202C}{ }

\def\BibTeX{{\rm B\kern-.05em{\sc i\kern-.025em b}\kern-.08em
    T\kern-.1667em\lower.7ex\hbox{E}\kern-.125emX}}
\begin{document}

\title{Blind Localization of Room Reflections with Application to Spatial Audio
}

\author{\IEEEauthorblockN{1\textsuperscript{st} Yogev Hadadi}
\IEEEauthorblockA{\textit{Electrical and Computer Engineering} \\
\textit{Ben-Gurion University of the Negev}\\
Beer-Sheva 84105, Israel \\
yogevhad@post.bgu.ac.il}
\and
\IEEEauthorblockN{2\textsuperscript{nd} Vladimir Tourbabin}
\IEEEauthorblockA{\textit{Reality Labs Research @ Meta} \\
Redmond, WA USA \\
vtourbabin@meta.com}
\and
\IEEEauthorblockN{3\textsuperscript{rd} Paul Calamia}
\IEEEauthorblockA{\textit{Reality Labs Research @ Meta} \\
Redmond, WA USA \\
‫pcalamia@gmail.com‬}
\and
\IEEEauthorblockN{4\textsuperscript{th} Boaz Rafaely}
\IEEEauthorblockA{\textit{Electrical and Computer Engineering} \\
\textit{Ben-Gurion University of the Negev}\\
Beer-Sheva 84105, Israel \\
br@bgu.ac.il}
} 

\maketitle

\begin{abstract}
Blind estimation of early room reflections, without knowledge of the room impulse response, holds substantial value. The FF-PHALCOR (Frequency Focusing PHase ALigned CORrelation) method was recently developed for this objective, extending the original PHALCOR method from spherical to arbitrary arrays. However, previous studies only compared the two methods under limited conditions without presenting a comprehensive performance analysis. This study presents an advance by evaluating the algorithm's performance in a wider range of conditions. Additionally, performance in terms of perception is investigated through a listening test. This test involves synthesizing room impulse responses from known room acoustics parameters and replacing the early reflections with the estimated ones. The importance of the estimated reflections for spatial perception is demonstrated through this test. 
\end{abstract}


\section{Introduction}

Numerous signal processing tasks such as optimal beamforming \cite{javed2016spherical}, room geometry inference \cite{mabande2013room,10096005}, source separation \cite{vincent2014blind}, speech enhancement and dereverberation \cite{kowalczyk2017extraction}, \cite{peled2010method}, may benefit from knowledge about early room reflections. Furthermore, early reflections affect sound perception by enhancing speech intelligibility, creating a feeling of listener envelopment, and enabling the evaluation of source features such as width, loudness, and distance \cite{catic2015role}, \cite{vorlander2007auralization}. Hence, methods for the blind estimation of early reflections have potential in applications of spatial audio signal processing\cite{pulkki2018parametric,coleman2017object,rafaely2022spatial}.

Methods for the blind estimation of room reflections have been previously suggested. These include beamforming \cite{mabande2013room}, MUSIC and ESPRIT \cite{sun2012localization, jo2019robust, ciuonzo2015performance, ciuonzo2017time}. However, they presented an estimation of only a few reflections due to the inherent limited spatial resolution of the array\cite{kuttruff2016room} and the density of reflections in space and time. Recently, a method was proposed that overcame the limitations of these previous methods. The method, named PHALCOR\cite{Shlomo2021Phalcor}, used a phase-aligned transform to decorrelate reflections along time and spatial decomposition to separate them spatially. Finally, utilizing time-space clustering, the method accurately estimated dozens of room reflections and outperformed previous methods. However, this method was only developed and investigated for spherical arrays using spherical harmonics signals. In follow-up work, a more generalized method named FF-PHALCOR \cite{hadadi2022towards} was proposed based on frequency focusing \cite{BeitOn2020Focusing}, extending the method to arbitrary arrays. This recent study demonstrated that the method achieved performance comparable to PHALCOR but explored only a limited set of conditions without presenting a comprehensive analysis.

This study offers a more comprehensive investigation of the FF-PHALCOR algorithm by evaluating the performance in a wider range of conditions. Furthermore, the analysis extends to spatial perception, where a listening test is conducted to examine the impact of estimating the Directions of Arrival (DOAs) of early reflections on the reproduction of room impulse responses (RIR) compared to synthesis based on a few room acoustics parameters. While a spherical array is still used in this paper, the algorithm operates directly on microphone signals and not in the spherical harmonics domain, therefore, moving towards operating with arbitrary arrays. 

The paper is structured as follows: Section \ref{Model} describes the system model, and Section \ref{Algorithm} describes the proposed algorithm. Next, Section \ref{Simulation} presents the results of a simulation study conducted to evaluate performance under various acoustic conditions. Section \ref{Listening} presents the results of the listening test. Finally, the paper is concluded in Section \ref{Conclusions}.

\section{Signal Model} \label{Model}
The acoustic scenario is a room consisting of a single source and a microphone array with $Q$ microphones. The speech signal in the frequency domain ($f$) is denoted by $\psi(f)$. The presence of $K$ early reflections is assumed, considered attenuated and delayed replicas of the anechoic source signal \cite{allen1979image}. The $k^{th}$ reflection is represented as a distinct source, $\emph{s}_k(f)$, with a direction of arrival (DOA) $\Omega_k$, delay $\tau_k$, and attenuation factor $\alpha_k$. $k=0$ represents the direct sound with the DOA $\Omega_0$, and normalized parameters such that $\tau_0=0$ and $\alpha_0=1$. Then:
\begin{equation}
    s_k(f)=\alpha_k e^{-i2\pi f\tau_k}\psi(f)
\end{equation}
  The vector $\mathbf{s}(f):= [s_0(f),s_1(f),\dots,s_K(f)]^T$ represents the concatenation of the direct sound and its $K$ early reflections arranged in an ascending order of delay. The pressure captured at the microphone array is represented by $\mathbf{p}(f):= [p_1(f),p_2(f),\dots,p_Q(f)]^T$ where each element corresponds to the pressure captured by a single microphone, resulting in the following array equation:

\begin{equation}\label{eq:pressure_on_mic}
   \mathbf{p}(f) = \mathbf{H}(f,\mathbf{\Omega})\mathbf{s}(f)+\mathbf{n}(f)
\end{equation}
$\mathbf{n}(f)$ includes the captured noise and the late reverberation, and the steering matrix $\mathbf{H}(f,\mathbf{\Omega})$ is constructed as follows:
\begin{equation} \label{eq:H_definition}
    \mathbf{H}(f,\mathbf{\Omega}) := [\mathbf{h}(f,\Omega_0),\dots,\mathbf{h}(f,\Omega_K)]
\end{equation}
where each $\mathbf{h}(f,\Omega_k)$ represents the steering vector in a free field of the $k^{th}$ reflection. Then, a frequency-independent steering matrix is obtained by employing frequency-focusing over the signal to apply the Phase-Alignment transform.

\section{FF-PHALCOR Algorithm} \label{Algorithm}

The first stage of the algorithm is frequency focusing, required to equalize the steering vectors within a selected band (with fixed bandwidth denoted by $BW$) centered around the center frequency $f_0$. Then, for each frequency $f$ within the band, a focusing matrix $\mathbf{T}(f, f_0)$ is computed by solving\cite{BeitOn2020Focusing,hadadi2022towards}:

\begin{equation}\label{eq:define_T}
\mathbf{T}(f,f_0)\mathbf{H}(f,\mathbf{\Omega}) = \mathbf{H}(f_0,\mathbf{\Omega})    
\end{equation} 
where $\mathbf{H}(f_0,\mathbf{\Omega})$ and  $\mathbf{H}(f,\mathbf{\Omega})$ are constructed using equation \eqref{eq:H_definition}. Applying focusing to equation \eqref{eq:pressure_on_mic} leads to:

\begin{equation}\label{eq:p_tilde_2}
    \tilde{\mathbf{p}}(f) = \mathbf{T}(f,f_0)\mathbf{p}(f) = \mathbf{H}(f_0,\mathbf{\Omega})\mathbf{s}(f) + \mathbf{T}(f,f_0)\mathbf{n}(f)
\end{equation}
After applying focusing, the spatial correlation matrix is computed:

\begin{equation}\label{eq:define_R_p_tilde}
    \mathbf{R}(f) = \mathbb{E}[\tilde{\mathbf{p}}(f)\tilde{\mathbf{p}}(f)^H]
\end{equation}
This matrix is usually dense in space and time, and to make it more sparse to allow improved separation of reflections, the Phase-Alignment transform is introduced as follows: 

\begin{equation}\label{eq:define_R_bar}
    \overline{\mathbf{R}}(\tau,f) := \sum_{j=0}^{J_f-1} \omega_j  \mathbf{R}(f_j)e^{i2\pi \tau j \Delta f}
\end{equation}
The parameter $J_f$ represents the total frequency points, $\Delta f$ is the frequency resolution, and $f_j = f+j\Delta f$. The weights $\omega_0,\dots,\omega_{J_f-1}$ are non-negative and inversely proportional to $tr( \mathbf{R}(f_j))$. In the original work \cite{Shlomo2021Phalcor}, it was shown that taking $\tau = \tau_k-\tau_{k'}$, with $k'=0$ (the direct sound) enhances entries in $\overline{\mathbf{R}}(\tau,f)$ which corresponds to reflections with delay $\tau_k$. Truncating its singular-value decomposition (SVD) provides a rank-1 approximation of $\overline{\mathbf{R}}(\tau)$, denoted by $\overline{\mathbf{R}}_1(\tau)$.

\begin{equation}\label{eq:svd}
    \overline{\mathbf{R}}_1(\tau) = \sigma_{\tau}\mathbf{u}_{\tau}\mathbf{v}_{\tau}^H
\end{equation}
For each $\tau$, the first singular value is denoted as $\sigma_{\tau}$, and the left and right singular vectors are represented by $\mathbf{u}_{\tau}$ and $\mathbf{v}_{\tau}$, respectively. It is shown that the former represents a superposition of the steering vectors of reflection arriving at a similar delay, while the latter represents the steering vector of the direct sound. The final stage of the algorithm is detecting and clustering of estimated delays and DOAs, performed using the DBSCAN algorithm \cite{ester1996density}, where identified clusters represent reflections. Further algorithm details can be found in \cite{Shlomo2021Phalcor,hadadi2022towards}.

\section{Simulation Study}\label{Simulation}

This study conducted a simulation to assess performance under various acoustic conditions. In the following section, a listening test was carried out using data from one of the simulated rooms.

\subsection{Simulation Setup}

The simulation setup involved shoe-box rooms with dimensions specified in Table \ref{tb:Rooms_table}, a speaker modeled as a point source, and a rigid spherical microphone array with a radius of $4.2\,$cm consisting of $32\,$microphones, similar to the Eigenmike \cite{acoustics2013em32}. The distance between the source and the array is also detailed in the table. The Room Impulse Response (RIR) from the speaker to the array was generated using the image method \cite{allen1979image}, while the speech signal used in the simulation was a $2.5\,$second sample from the TSP Speech Database \cite{kabal2002tsp}, recorded at a sampling frequency of $48\,$kHz. The microphone signals were generated using a sound field model of order $N=8$ in the spherical harmonics domain, generated directly from the image source data \cite{allen1979image, rafaely2015fundamentals}. The Direct to Reverberant Ratio (DRR) is calculated from the RIR as in Eq (1) in \cite{calamia2020blind}, $T_{60}$ and critical distance are obtained by the diffuse model using Sabine formula \cite{sabine1994collected, kuttruff2012room,sabine1900reverberation, kinsler2000fundamentals} for each simulated room and scene, all presented in the table.

\begin{table}[bp]
\caption{Dimensions, $T_{60}$, Critical distance ($R_c$), source-array Distance and DRR of the simulated rooms and acoustic scenes}
\begin{center}
\begin{tabular}{|c|c|c|c|c|c|}
\hline
\textbf{Room} & \textbf{Dim $\left[m\right]^3$} &  \textbf{$\mathbf{T_{60} [s]} $} &\textbf{ $R_c$ [m]} & \textbf{Distance [m]} & \textbf{DRR [dB]}\\
\hline
\textbf{1} & [8,6,4] & 0.413 & 1.22 & 4.03 & -11.5\\
\hline
\textbf{2} & [7,5,3] & 0.186 & 1.34 & 3.00 & -4.0  \\
\hline
\textbf{3} & [9,7,4] & 0.841 & 0.98 & 1.28 & -5.2  \\
\hline
\textbf{4} & [11,7,4] & 0.876 & 1.06 & 1.97 & -7.57  \\
\hline
\textbf{5} & [13,8,5] & 1.055 & 1.26 & 4.90 & -13.6  \\
\hline

\end{tabular}
\label{tb:Rooms_table}
\end{center}
\end{table}


\subsection{Methodology}

First, a short-time Fourier transform (STFT) is computed for the microphone signals $p_q$, using the Hanning window with a size of $150\,$ms, $f_s=48\,$kHz sampling frequency, and an overlap of $75\%$. An operating frequency range of $[500,5000]\,$Hz was selected, with a bandwidth of $2\,$kHz for focusing, $J_f=8$ frequency bins was utilized for equation (\ref{eq:define_R_bar}) \cite{Shlomo2021Phalcor}. Focusing is applied using a simulated steering matrix constructed with 900 DOAs sampled by the Fliege-Maier method \cite{fliege1996two}. Finally the hyper-parameters as defined in the original paper\cite{Shlomo2021Phalcor} were selected as follows: $\rho_{min}=0.9$, $\epsilon_u = 0.63$, $\Omega_{th} = 10^{\circ}$, $S_{max} = 3$, $\gamma_{\Omega} = 8^{\circ}$ and $\gamma_{\tau} = 0.3\,$ms, and the density threshold is $0.15$. 

For the performance evaluation, a reflection is classified as a true positive if it matches a true reflection (obtained using the image method) within specified tolerances set to $0.5\,$ms in delay, and $15^{\circ}$ in DOA. The Probability of Detection (PD) is defined as follows:

\begin{equation}\label{eq:PD}
    PD := \frac{\#\;true\;positive\;detections}{\#reflections\;in\;the\;ground\;truth}
\end{equation}

The Probability of False Alarm (PFA) is defined as follows:

\begin{equation}\label{eq:PFA}
    PFA := \frac{\#\;false\;positive\;detections}{\#detected\;reflections}
\end{equation}

\subsection{Results}

\begin{figure}[htbp]
    \begin{center}
         \centering
         \includegraphics[width=0.95\columnwidth]{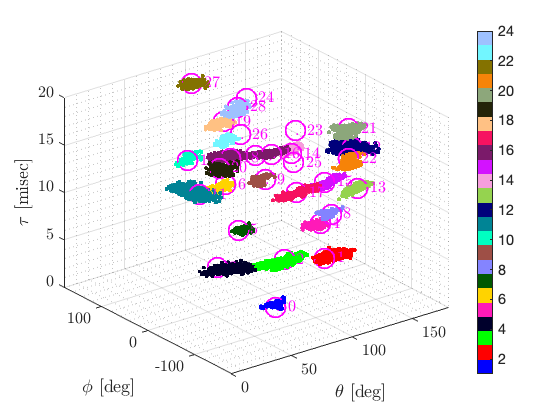}
         \label{fig:anm5}     
  
        \caption{Clustering of DOA, delay obtained in room1. $\tau$ [ms] is the delay, $\theta$,$\phi$ [deg] are the elevation and the azimuth. The rings are the ground truths.}
        \label{fig:room1}
    \end{center}

\end{figure}

Fig. \ref{fig:room1} presents the clustering of the algorithm in the first room. The axes are the elevation angle $\theta$, the azimuth angle $\phi$ both in degrees, and the delay $\tau$ in milliseconds. The pink circles are the ground truth obtained from the image method. The colored regions are the results of the clustering algorithm. A more detailed analysis of this result revealed that the algorithm successfully detected the direct sound and 23 out of 28 reflections with $PD = 82\%$ and $PFA=0$. As partially seen in this figure, the algorithm better estimates sparser reflections compared to denser reflections as the clustering algorithm may mistakenly group multiple reflections.

Table \ref{tb:results} displays the algorithm's performance under various conditions. The first column is the room number, corresponding to Table \ref{tb:Rooms_table}, and the second presents the number of reflections within $20\,$ms of the RIR. The third column presents the number of reflections the algorithm detected, including false alarms. Finally, the last columns present the PD and PFA calculated as in equations (\eqref{eq:PD}) and (\eqref{eq:PFA}).
It can be seen that overall the algorithm performs well and detects about $18-29$ reflections in each room $(1,2,5)$ and $8-9$ in $3,4$.

\begin{table}[htbp]
\caption{Simulations results, where $\#$Reflections is the number of reflections within the first 20ms of the RIR (including the direct sound), $\#$Detections (including false-alarms). PD, PFA are also presented}
\begin{center}
\begin{tabular}{|c|c|c|c|c|}
\hline
\textbf{Room} & $\#$Reflections & $\#$Detections** & PD & PFA \\
\cline{1-5} 
\textbf{1} & 29 & 24 & $82\%$ & 0\\
\hline
\textbf{2} & 38 & 29 & $52\%$ & $24\%$ \\
\hline
\textbf{3} & 9 & 9 & $100\%$ & 0 \\
\hline
\textbf{4} & 8 & 8 & $100\%$ & 0 \\
\hline
\textbf{5} & 16 & 18 & $94\%$ & $17\%$ \\
\hline
\end{tabular}
\label{tb:results}
\end{center}
\end{table}

\section{Listening Test}\label{Listening}

As demonstrated in the preceding section, the algorithm successfully detects early reflections. It has been established \cite{kitic2023blind,rafaely2022spatial} that early reflections play an important role in spatial perception. To evaluate the application of FF-PHALCOR in the context of spatial perception, a listening test was conducted based on the MUSHRA test (Multiple Stimuli with Hidden Reference and Anchor) \cite{series2014method}.

\subsection{Setup}

The setup comprises a shoe-box room with the same dimensions as room 1 in Table \ref{tb:Rooms_table}. It includes a speaker located at $[5.5, 1.2, 1.7]$ and a $32$-channel spherical microphone array located at $[2.5, 3.9, 1.7]$, with $T_{60} = 0.424\,$seconds, and DRR $= -11.5\,$dB. The signal is a $2.5\,$second segment extracted from the TSP Speech Database \cite{kabal2002tsp}. The head-related transfer function (HRTF) used in the experiment is derived from the Neumann KU-$100$\cite{bernschutz2013spherical}.

\subsection{Methodology}

The test comprises three signals: the reference signal, a synthesized anchor signal, and the estimated signal. The reference signal is generated by applying the image method to compute the room impulse response with an order of $N=40$.

The synthesized anchor signal was generated using known room acoustics parameters: the room volume $V$, $T_{60}$, DRR, and DOA of the direct sound. In this case, the room volume is $V=192\,m^3$, and the DOA of the direct sound is given as $(90,-42.2)\,$degrees, the elevation and azimuth angle in the range of $[0,\pi]\times[-\pi,\pi)$, respectively. Next, the expected cumulative number of reflections was computed using a model based on the image method \cite{allen1979image}, and this estimated number was used to generate reflection delays in 5ms frames randomly. The amplitude of the reflections was selected to fit a decay curve to achieve the desired $T_{60}$ value  as calculated using the Schroeder integral \cite{schroeder1965new}. Reflections' DOAs were randomly selected from a uniform distribution over the sphere. The first reflection is considered as the direct sound with a known DOA, and its amplitude is calculated to maintain the DRR value.

For the estimated signal, the initial 20ms of the synthesized RIR described above is replaced with the reflections estimated by the algorithm, keeping the amplitudes the same as in the synthesized. Subsequently, all signals are convolved with the head-related impulse response (HRIR) to create a binaural room impulse responses (BRIR)\cite{rafaely2010interaural}. Then, the speech signal is convolved with the BRIR to generate the signals. Finally, the synthetic and estimated signals were equalized in power to equal the reference. As a final stage, headphone equalization was applied. 

A total of 12 participants with normal hearing took part in the listening test. The experiment consisted of a single MUSHRA screen, where all three signals were played back using the audio player in MATLAB (MATLAB R2022b). Participants were instructed to rate the overall quality on a scale from 0 to 100. Before rating, participants were given a training task to ensure understanding of the instructions and familiarize themselves with the signals.

\subsection{Results}

\begin{figure}[h]
    \begin{center}
         \centering
         \includegraphics[width=0.95\columnwidth]{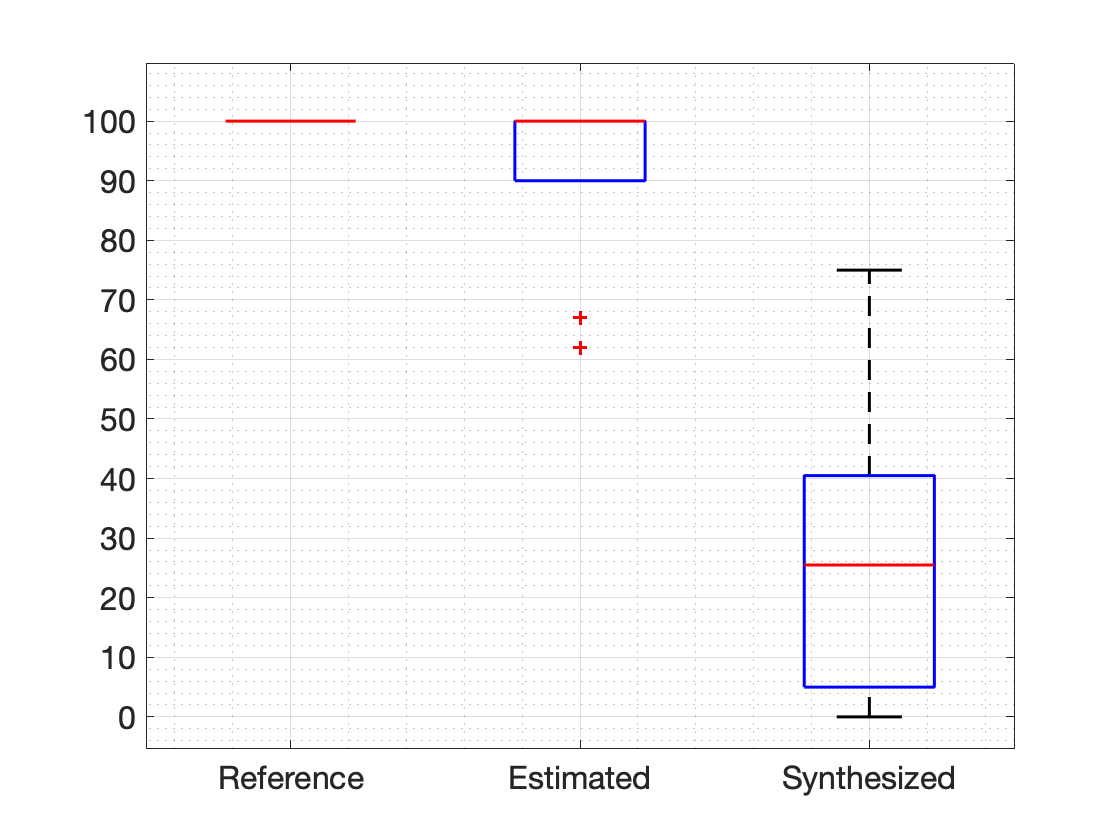}
  
        \caption{The results of the overall quality test, in a box plot visualization. The red line represents the median, the edges represent the 25th and 75th percentiles. The whiskers show the range of values excluding outliers. Outliers are marked with a red "+" symbol.}
        \label{fig:listening_test_results}
    \end{center}

\end{figure}

Fig. \ref{fig:listening_test_results} displays the test results conducted in this study. The three Boxes represent the evaluated signals, and the y-axis represents the similarity to the reference signal. Furthermore, a Repeated Measures ANalysis of VAriance (RMANOVA)\cite{keselman2001analysis} has been performed. In this test, one within-subject variables was analyzed, i.e the reproduction method. The method variable exhibited a statistically significant main effect, as revealed by the RM-ANOVA analysis, a $F(2,22) = 76.54$, $p<0.001$ and $\eta_p^2=0.874$. Given the statistically significant main effect observed for the method variable, a post-hoc test was conducted with Bonferroni correction to investigate this result further. Pairwise comparisons were made between the estimated marginal means. The reference and the estimated signals were not statistically significant, with a mean difference of $7.583$ and $p=0.24$. However, the reference and the synthesized signals and the estimated and the synthesized signals are statistically significant with mean differences of $72.083,64.5$ respectively, and $p<0.001$. This indicates that the algorithm successfully enhanced the spatial perception by incorporating the knowledge of early room reflections.

These findings highlight the effectiveness of the FF-PHALCOR algorithm in improving spatial perception by incorporating estimated early room reflections into the reproduced signals.

\section{Conclusions}\label{Conclusions}
This paper showcases the effectiveness of the proposed algorithm in detecting a significant number of reflections. Moreover, it demonstrates the impact of the algorithm's estimation of early reflections on sound perception. These findings represent an advancements in the extension of the PHALCOR method to arbitrary arrays.

For future work, it is recommended to study the algorithm's performance over non-spherical arrays, and to deepen the impact of the algorithm on the spatial perception of reproduced signals.

\bibliographystyle{IEEEtran}
\bibliography{refs.bib}

\end{document}